\documentclass{article}
\usepackage[T1]{fontenc}
\usepackage{a4wide}
\usepackage{graphics}

\makeatletter

\providecommand{\LyX}{L\kern-.1667em\lower.25em\hbox{Y}\kern-.125emX\@}

\newcommand{\lyxrightaddress}[1]{
  \par {\raggedleft \begin{tabular}{l}\ignorespaces
  #1
  \end{tabular}
  \vspace{1.4em}
  \par}
}

\usepackage[T1]{fontenc}

\makeatletter

\usepackage[T1]{fontenc}

\makeatletter

\usepackage[T1]{fontenc}

\makeatletter

\usepackage[T1]{fontenc}

\makeatletter

\usepackage[T1]{fontenc}


\makeatother
\makeatother
\makeatother

\makeatother

\begin{document}

\title{Charged Particle Fluctuations as a Signal of the Dynamics of Heavy Ion Processes}

\author{Fritz W. Bopp and Johannes Ranft \\
   \\
   Universität Siegen, Fachbereich Physik, \\
 D--57068 Siegen, Germany }

\maketitle
\vspace*{-8cm}

\lyxrightaddress{SI-00-9}

\vspace*{+7cm}

\begin{abstract}
Comparing proposed quantities to analyze charged particle fluctuations in heavy
ion experiments we find the dispersion of the charges in a central rapidity
box as best suited. Various energies and different nuclear sizes were considered
in an explicit Dual-Parton-Model calculation using the DPMJET code. A definite
deviation from predictions of recently considered statistical models was obtained.
Hence the charged particle fluctuations should provide a clear signal of the
dynamics of heavy ion processes. They should allow to directly measure the degree
of thermalization in a quantitative way.
\end{abstract}

\section{Introduction}

In the analysis of the hadronic multi-particle production (for a recent review
see \cite{jacob99}) a key observation has been the local compensation of charge\cite{idschok73}.
The charge fluctuations connected to the soft hadronic part of the reactions
were found to involve only a restricted rapidity range. This observation limited
the applicability of statistical models to rather local fluctuations (see e.g.
\cite{ranft75} where the charge fluctuation between the forward and backward
hemisphere\cite{chouyang73} was discussed). 

In heavy ion scattering charge flow measurements should be analogously decisive.
It is a central question of an unbiased analysis whether the charges are distributed
just randomly or whether there is some of the dynamics left influencing the
flow of quantum numbers. This is not an impractical conjecture. In heavy ion
experiments the charge distribution of particle contained in a central box with
a given rapidity range \( [-Y_{\mathrm{max}.},+Y_{\mathrm{max}.}] \) can be
measured and the dispersion of this distribution:
\begin{equation}
\label{eq-1}
<\delta Q^{2}>=<(Q-<Q>)^{2}>
\end{equation}
can be obtained to sufficient accuracy. For sufficiently large gaps this quantity
also reflects the long range charge flow.

It was proposed to use this quantity to distinguish between particles emerging
from a equilibrized quark-gluon gas or from a equilibrized hadron gas\cite{heinz00,jeon00,jeon99}.
In a hadron gas each particle species in the box is taken essentially poissonian.
In a central region where the average charge flow can be ignored, one obtains
a simple relation for particles like pions with charges \( 0 \) and \( \pm 1 \)
:

\begin{equation}
\label{eq-2}
<\delta Q^{2}>=<N_{\mathrm{charged}}>.
\end{equation}
 It is argued in the cited papers that this relation would change in a quark
gluon gas to:
\begin{equation}
\label{eq-3}
<\delta Q^{2}>=\sum _{i}q_{i}^{2}<N_{i}>=0.19<N_{\mathrm{charged}}>
\end{equation}
where \( q_{i} \) are the charges of the various quark species and where again
a central region is considered. The coefficient on the right was calculated\cite{jeon00}
for a two flavor plasma in a thermodynamical consideration which predicts various
quark and gluon contributions with suitable assumptions. A largely empirical
final charged multiplicity \( N_{\mathrm{charged}}=\frac{2}{3}(N_{\mathrm{glue}}+1.2N_{\mathrm{quark}}+1.2N_{\mathrm{antiquark}}) \)
was used. 

It should be pointed out that the estimate is not trivial. In the considered
\( m_{\mathrm{quark}}=0 \) theory the observable \( Q \) is not infrared safe
and makes no sense. A way to make it well defined\cite{bopp75} is to consider
the quantity:
\begin{equation}
\label{eq-4}
Q_{\mathrm{quark}}\rightarrow \, \widetilde{Q}_{\mathrm{quark}}=Q_{\mathrm{quark}}-<Q_{\mathrm{quark}}>
\end{equation}
which avoids the influence of extra sea quarks. Such a correction is also needed
to have an observable which can be expected to survive hadronization. Numerically
the effect is not very big and the problems can be ignored if only a rough description
is sufficient.

There are a number of sources of systematic errors in the above comparison.
The result strongly depends on what one chooses as primordial and secondary
particles. Considering these uncertainties we follow Fia{\l }kowski's conclusion\cite{fialkowski00}
that a clear cut distinction between the hadron- and the quark gluon gas is
rather unlikely. This does not eliminate the interest in the dispersion. The
hadron gas model is anyhow no optimal reference point to compare with.

In the next section we discuss various possible measures to observe such fluctuations.
We favor the dispersion of the charge transfer. Using an explicit Dual Parton
model calculation we observe a clear distinction between models with local compensation
of charge and equilibrium approaches. In section 3 a simple interpretation of
the dispersion in terms of quark lines is outlined. This suggests to compare
the dispersion to the particle density as it is done in section 4.

\section{Various Measures for Fluctuations in the Charge Distribution }

\vspace{0.3cm}
For the analysis of the charge structure several quantities were discussed in
the recent literature. It was proposed to look at the particles within a suitable
kinematic region and to measure just the mean standard deviation of the ratio
\( R \) of positive to negative particles: 
\begin{equation}
\label{eq-5}
<\delta R^{2}>=<\left( \frac{N_{+}}{N_{-}}\: -<\frac{N_{+}}{N_{-}}>\right) ^{2}>
\end{equation}
or the quantity \( F \):
\begin{equation}
\label{eq-6}
<\delta F^{2}>=<\left( \frac{Q}{N_{\mathrm{charged}}}-<\frac{Q}{N_{\mathrm{charged}}}>\right) ^{2}>
\end{equation}
where \( Q=N_{+}-N_{-} \). 

We consider them not attractive. The quantities are not suitable for small intervals,
as there are actually undefined in a certain region. They are less clean than
the dispersion, \( <\delta Q> \), as they are not exclusively connected to
the flavor structure and as they mix up charge and density fluctuations. 

For large nuclei at high energies this is not a problem as the density fluctuations
are small and all three quantities are connected by the following relations\cite{jeon00}:
\begin{equation}
\label{eq-7}
<N_{\mathrm{charged}}><\delta R^{2}>=4\: <N_{\mathrm{charged}}><\delta F^{2}>=4\cdot \frac{<\delta Q^{2}>}{<N_{\mathrm{charged}}>}.
\end{equation}
 To examine the range where these relations hold, all three quantities were
calculated in the Dual Parton model implementation DPMJET\cite{DPMJET} . For
the most central \( 5\% \) Pb-Pb scattering at LHC energies (\( \sqrt{s}=6000 \)
A GeV) there is indeed a perfect agreement between all three quantities as shown
in figure 1.  For the most central \( 5\% \) S-S scattering at RHIC energies
(\( \sqrt{s}=200 \) A GeV) the agreement is no longer as good. For the minimum
bias S-S scattering at RHIC energies the agreement is lost and the first two
expressions behave rather erratic. The same erratic behavior is seen for the
proton-proton case which is shown figure 2. As any conclusion will strongly
depend on a comparison of central processes with minimum bias events, we advocate
to stick to the dispersion of the net charge distribution \( <\delta Q^{2}> \). 

We observed no significant difference between rapidity and pseudo-rapidity boxes.
\begin{figure*}
{\par\centering \resizebox*{!}{0.4\textheight}{\includegraphics{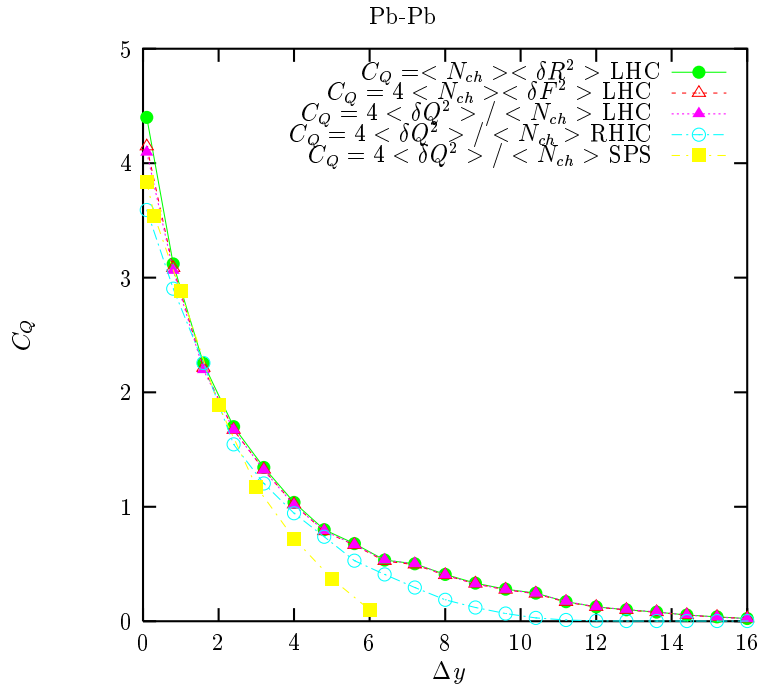}} \par}

\caption{Charge fluctuations for the most central \protect\( 5\%\protect \) Pb-Pb scattering
at LHC energies (\protect\( \sqrt{s}=6000\protect \) A GeV). Also shown are
dispersions for RHIC (\protect\( \sqrt{s}=200\protect \) A GeV) and SPS (\protect\( E_{}ab\protect \)
A GeV) energies.}
\end{figure*}
\begin{figure}
{\par\centering \resizebox*{!}{0.4\textheight}{\includegraphics{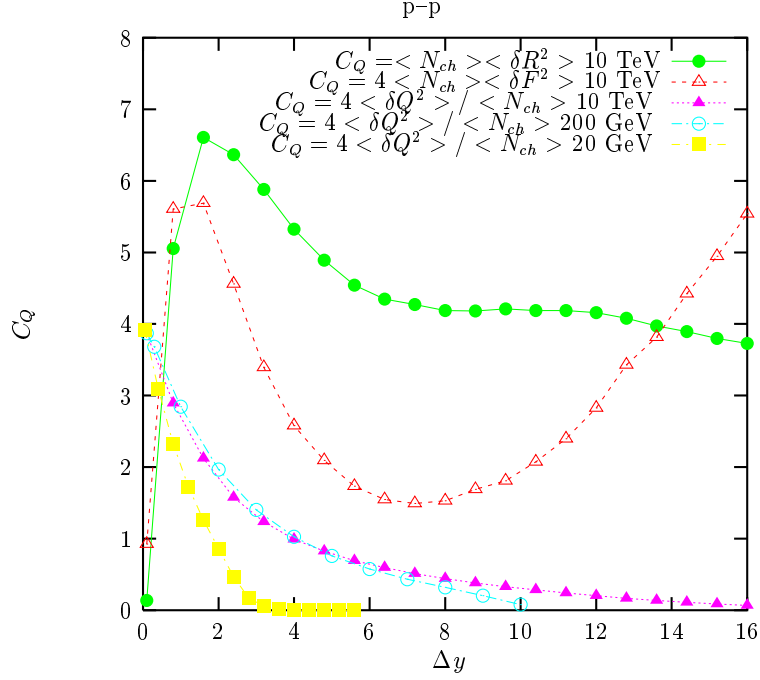}} \par}

\caption{Charge fluctuations for minimum bias pp scattering at SPS, RHIC and LHC energies}
\end{figure}

\section{A Simple Relation between the Quark Line Structure and Fluctuations in the
Charge Flow}

To visualize the meaning of charge flow measurements it is helpful to introduce
a general factorization hypothesis. It is not exact but it is widely expected
to hold to good accuracy. It postulates that the flavor structure of an arbitrary
amplitude can be described simply by an overall factor, in which the contribution
from individual quark lines factorize\footnote{
The hypothesis is based on the exchange degeneracy of octet and singlet Regge
trajectories effectively changing the \( SU(N_{\mathrm{effective}}) \) flavor
symmetry to an \( U(N_{\mathrm{effective}}) \) symmetry in which this relation
is exactly valid. One of the important corrections to the hypothesis originates
in the special behaviour of the masses of the lowest lying mesons of the trajectories,
which is especially significant in the pseudo-scalar sector between the \( \pi _{0} \)
and the \( \eta  \) meson. That one of the two neutral states is sometimes
suppressed, introduces a mild anticorrelation between neigbouring flavors, which
can be ignored for our consideration centered at long range charge transfers.

If a higher accuracy is desired the hypothesis can be restricted to primordial
particles generated at a high ``temperature'' and ``secondary'' charges
produced during the decay of primordial particles can be considered extra. With
simple assumptions their contribution can be related to the corresponding particle
spectrum. If all charged particles were secondaries the dispersion of the charge
transfer across an arbitrary rapidity boundary would be given by the Quigg-Thomas
relation\cite{quigg73,quigg75,baier74} \( <\delta Q^{2}(y)>=\sigma \frac{1}{2}\rho _{\mathrm{charged}}(y) \)
where \( \sigma =1 \) if widening and narrowing effects balance. 
}. 

The hypothesis can be used to obtain the following generalization of the Quigg-Thomas
relation\cite{baier74,aurenche77,bopp78}. This generalization states that the
correlation of the charges \( Q(y_{1}) \) and \( Q(y_{2}) \), which are exchanged
during the scattering process across two separate kinematic boundaries, is just:
\begin{equation}
\label{eq-8}
<\left\{ Q(y_{1})-<Q(y_{1})>\right\} \left\{ Q(y_{2})-<Q(y_{2})>\right\} >=n_{\mathrm{common}\: \mathrm{lines}}<(q-<q>)^{2}>.
\end{equation}
where \( n_{\mathrm{common}\: \mathrm{lines}} \) counts the number of quark
lines intersecting both borders and \( q \) is the charge of the quark on such
a line. Depending on the flavor distribution average values \( <(q-<q>)^{2}=0.22\cdots 0.25 \)
are obtained. 

Most observables of charge fluctuations depend on this basic correlation. The
fluctuation of the charges within a \( [-Y_{\mathrm{max}.},+Y_{\mathrm{max}.}] \)
box discussed above contains a combination of three such correlations: 
\begin{equation}
\label{eq-9}
<\delta Q^{2}>\: =\: <\delta Q(y_{1})^{2}>+<\delta Q(y_{2})^{2}>-\: 2<\delta Q(y_{1})\cdot \delta Q(y_{2})>.
\end{equation}
 Using (\ref{eq-8}) the dispersion of the charges in a box subtracts to:
\begin{equation}
\label{eq-10}
<\delta Q[\mathrm{box}]^{2}>=n_{\mathrm{lines}\: \mathrm{entering}\: \mathrm{box}}<(q-<q>)^{2}
\end{equation}
where \( n_{\mathrm{lines}\: \mathrm{entering}\: \mathrm{box}} \) is the number
of quark lines entering the box.

\section{Calculation of the Dispersion of the Charge Distribution within a Box}

Let us consider the prediction of a thermodynamic model in more detail. In the
thermodynamic limit with an infinite reservoir outside and a finite number of
quarks inside all quark lines will connect to the outside as shown in figure
3.
\begin{figure}
{\par\centering \resizebox*{!}{0.15\textheight}{\includegraphics{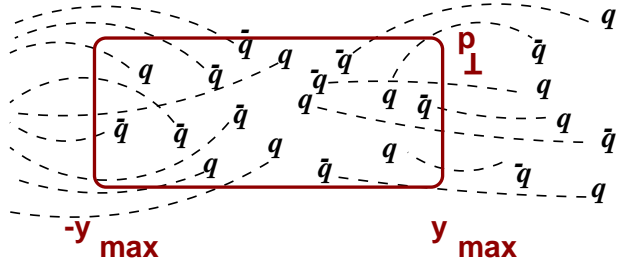}} \par}

\caption{Quark lines entering the box in the thermodynamic limit }
\end{figure}
The dispersion of the charge transfer is therefore proportional to the total
number of particles inside. In the hadron gas all particles contain two independent
quarks each contributing with roughly \( 1/4 \) yielding the estimate of (\ref{eq-2}).
For the quark gas only one quark originates in the thermalization and is taken
to be responsible for charge transfer yielding a considerable smaller result.
Obviously there are several refinements to this simple picture.

Let us consider the limit of a tiny box. Looking only at the first order one
trivially obtains:

\begin{equation}
\label{eq-11}
<\delta Q^{2}>/<N_{\mathrm{charged}}>=1
\end{equation}
which corresponds to the hadron gas value. 

If the box size increases to one or two units of rapidity on each side this
ratio will typically decrease, as most models contain a short range component
in the charge fluctuations usually attributed to secondary interactions. One
particular short range fluctuation might be caused by the hadronization of partons
of the quark gluon gas. The quark antiquark pair needed for the hadronization
is assumed to be short range so that for a box of a certain size one just obtains
the charge dispersion of the original partons. This is responsible for the reduction
of the fluctuation discussed above. The decreasing is however not very distinctive.
Common to many models are secondary interactions which involve decay processes
and comover interaction. In hadron hadron scattering processes such correlations
are known to play a significant role and there is no reason not to expect the
same for the heavy ion case.

After a box size passed the short range the decisive region starts. In all global
statistical models\cite{heinz00,jeon99,jeon00} the ratio will have to reach
now a flat value. Only for very large rapidity ranges charge conservation will
force the ratio to drop \( \left( \propto 1-y_{\mathrm{max}.}/Y_{\mathrm{kin}.\mathrm{max}.}\right)  \).
This is different in string models as it is illustrated in figures 1 and 2 .
The model calculation shows with its rapid fall off a manifestly different behavior.
It is a direct consequence of the local compensation of charge contained in
string models. The effect is illustrated in Figure~4 in which only quark lines
which contribute to the charge flow and which intersect the boundary are shown.
\begin{figure}
{\par\centering \resizebox*{!}{0.15\textheight}{\includegraphics{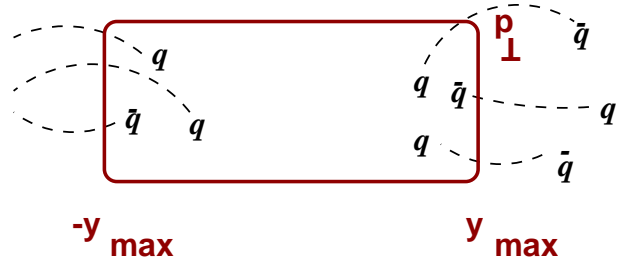}} \par}

\caption{Quark lines entering the box with local compensation of charge}
\end{figure}
The local compensation of charge allows now only a contribution of lines originating
around the boundaries. If the distance is larger that the range of charge compensation
the dispersion will no longer increase with the box size. The total contribution
will now be just proportional to the density of the particles around the boundaries
\begin{equation}
\label{eq-12}
<\delta Q^{2}>\, \propto \rho _{\mathrm{charged}}(y_{\mathrm{max}.}).
\end{equation}

This resulting scaling is illustrated in a comparison between both quantities
shown in Figure~5 for RHIC and LHC energies.
\begin{figure}
{\par\centering \resizebox*{!}{0.4\textheight}{\includegraphics{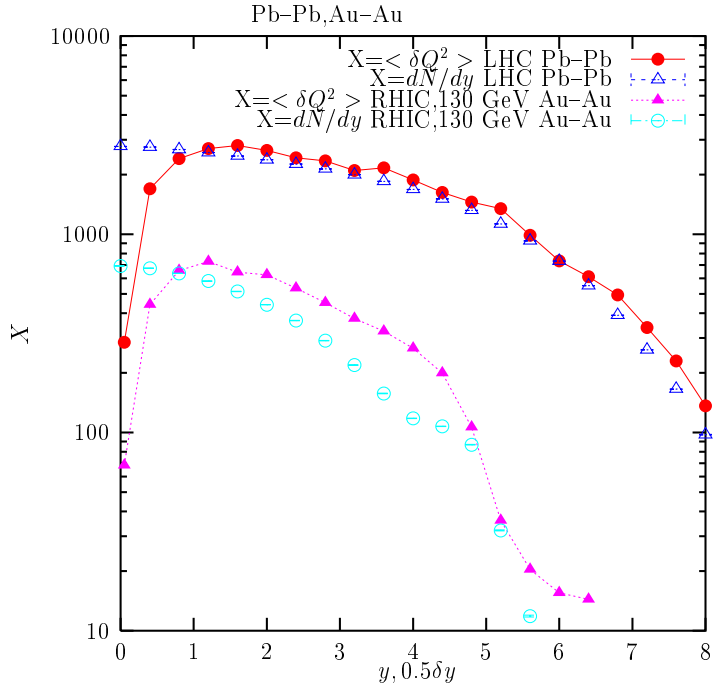}} \par}

{\par}

\caption{Comparison of the dispersion of the charge distribution with the density on
the boundary of the considered box for central gold gold resp. lead lead scattering
at RHIC and LHC energies.}
\end{figure}
 The agreement is comparable to the proton proton case shown in Figure 6.
\begin{figure}
{\par\centering \resizebox*{!}{0.4\textheight}{\includegraphics{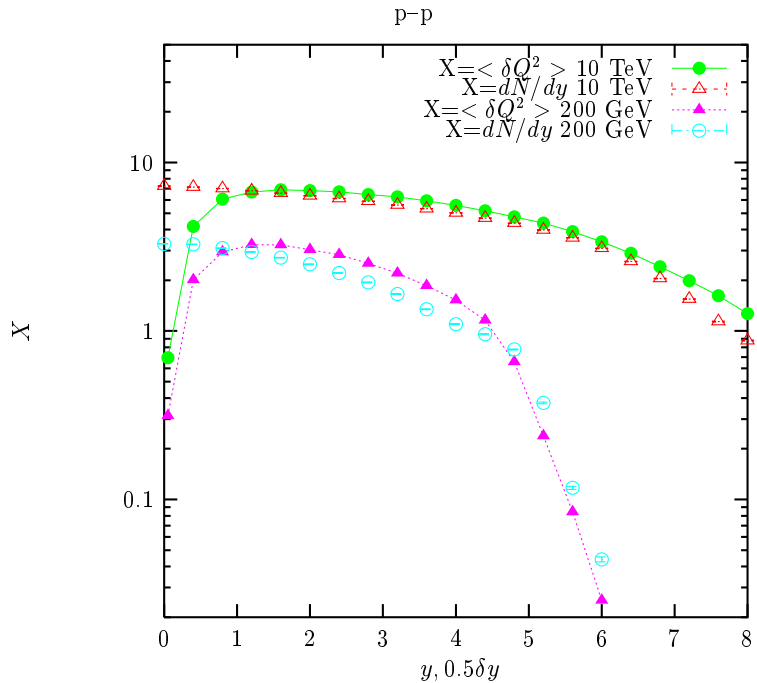}} \par}

\caption{Comparison of the dispersion of the charge distribution with the density on
the boundary of the considered box for proton proton scattering at RHIC and
LHC energies.}
\end{figure}
 The proportionality is expected to hold for a gap with roughly \( \frac{1}{2}\delta y>1 \)
. For smaller boxes some of the quark lines seen in the density do not contribute
as they intersect both boundaries. For large rapidity sizes there is a minor
increase from the leading charge flow \( Q_{L} \) originating in the incoming
particles. In a more careful consideration\cite{bopp78} one can subtract this
contribution
\begin{equation}
\label{eq-13}
<\delta Q^{2}>_{\mathrm{leading}\: \mathrm{charge}\: \mathrm{migration}}=<Q_{L}>(1-<Q_{L}>)
\end{equation}
 and concentrate truly on the fluctuation. 
\vspace{0.3cm}

The prediction for the proportionality factor for the case of mere short range
fluctuations would be roughly a factor one (see footnote 1). In string models
primordial particles are responsible for a longer range charge transfer coming
from the contributions of the quark resp. diquark fragmentation chains. Taking
everything together one obtains
\begin{equation}
\label{eq-14}
<\delta Q^{2}>=\sum _{\mathrm{left}+\mathrm{right}}\{\: n_{\mathrm{strings}}\cdot <(q-<q>)^{2}>+\: f_{\mathrm{secondary}}\: \sigma \: \frac{1}{2}\: \rho _{\mathrm{charged}}(y)\: \}
\end{equation}
 where \( n_{\mathrm{strings}} \) is the number of strings where \( f_{\mathrm{secondary}} \)
is the fraction of secondary to primary particles and where the width of the
local fluctuations \( \sigma  \) is roughly unity. In our Dual Parton Model
calculation we observe factor of roughly \( 1.2 \) between the density and
the dispersion. It means that most of the fluctuation originate in secondary
interaction and that the effective larger coefficient of the first term which
significantly rises the factor plays only a lesser role.

\section*{Conclusion}

In the paper we demonstrated that the dispersion of the charge distribution
in a central box of varying extend in rapidity is an extremely powerful measure.
Within the string model calculation the dispersion seen in relation to the spectra
shows no difference between simple proton proton scattering and central lead
lead scattering even though both quantities change roughly by a factor of 400. 

The dispersion allows to clearly distinguish between conventional string models
and thermal models. As most expected changes in the dynamic are somehow connected
to the onset of thermalization they will introduce more fluctuations. Even if
the truth should lie somewhere in between string models and quark gluon plasma
models it is therefore quite reasonable to hope that the position of the transition
can be determined in a quantitative way.

\section*{Acknowledgments}

F.W. Bopp acknowledges partial support from the INTAS grant 97-31696.

\end{document}